\documentclass[aps,pra,notitlepage,onecolumn, preprintnumbers,amsmath,showpacs, amssymb,10pt]{revtex4-1}

\usepackage{hhline}
\usepackage{booktabs}

\usepackage{times}

\usepackage[usenames,dvipsnames,table]{xcolor}

\usepackage{tikz}
\usetikzlibrary{arrows,shapes,positioning,shadows,backgrounds,fit}

\usepackage[english]{babel}
\usepackage{graphicx, dcolumn, braket, bm, color, amsmath, relsize, amsmath, dsfont, mathrsfs, empheq, verbatim, upgreek, etoolbox}
\usepackage[caption = false]{subfig}

\usepackage[colorlinks=true,linkcolor=blue,urlcolor=blue,citecolor=blue]{hyperref}
\usepackage{graphics}
\usepackage{bm}
\usepackage{color}
\usepackage{amscd}
\usepackage{amsfonts}
\usepackage{amssymb}
\usepackage{graphicx}

\newcommand{\R}{\mathbb{R}}
\newcommand{\C}{\mathbb{C}}
\newcommand{\sys}{\mathcal{S}}
\newcommand{\h}{\mathscr{H}}
\newcommand{\hsys}{\h_{\sys}}
\newcommand{\ha}{\h_{\mathcal{A}}}

\newcommand{\mean}[1]{\langle #1 \rangle}
\newcommand{\YUPPI}{\mathlarger{\mathlarger{\uppi}}}
\providecommand{\bgreek}[1]{\mbox{\boldmath$#1$}}
\newcommand{\vett}[1]{\mathbf{#1}}

\newcommand{\Id}{\mathds{1}}

\newcommand{\hbarra}{\h}
\providecommand{\bgreek}[1]{\mbox{\boldmath$#1$}}

\begin{document}

\newdimen\origiwspc%
  \newdimen\origiwstr%

\title{Unified treatment of the total angular momentum of single photons via generalized quantum observables}

\author{M. Motta}
\affiliation{Division of Chemistry and Chemical Engineering, California Institute of Technology, Pasadena, California 91125, USA}
\thanks{Both authors contributed equally to this work}

\author{G. Guarnieri}
\affiliation{Department of Optics, Palacky University, 17. listopadu 1192/12, 771 46 Olomouc, Czech Republic}
\thanks{Both authors contributed equally to this work}

\author{L. Lanz}
\affiliation{Dipartimento di Fisica, Universit\` a degli Studi di Milano, Via Celoria 16,
I-20133, Milano, Italy}

\author{B. Hiesmayr}
\affiliation{University of Vienna, Faculty of Physics, Boltzmanngasse 5, 1090 Vienna, Austria}


%
%

\begin{abstract}
In this paper we provide a consistent framework to address the notorious decomposition of the single-photon total angular momentum (TAM) into a spin (SAM) and an orbital (OAM) component. 
In particular, we find that the canonical SAM and OAM, generators of internal and spatial rotations respectively in the space of physical states, are mutually compatible but \textit{unsharp} quantum observables, therefore POVM (Positive Operator-Valued Measures) describe their joint measurements. 
On the other hand, a non-canonical SAM and OAM can be constructed that is mutually incompatible but represent \textit{sharp} quantum observables, thus PVM (Projector-Valued Measurements), reflecting their consistency with the transversality condition characterizing single–photon wavefunctions. 
We discuss the implementations on joint measurements for both decompositions and provide an explicit calculation of all these quantities for circularly polarized Gaussian single-photon states.
\end{abstract}

\date{\today}
\maketitle

\section{Introduction}

Understanding and manipulating the angular momentum of single photons is a central goal of modern
physics, due to theoretical, experimental and even technical implications.
On the experimental side, the angular momentum of light has been recently recognized as a novel
powerful resource for implementing quantum--information protocols~\cite{Allen1992,Molina2007,Nagali2009,Krenn2014,Hiesmayr2016,Hiesmayr2017}.
Moreover, experiments with light at the single--photon level have historically been at the forefront
of fundamental tests of quantum mechanics~\cite{Freedman1972,Ou1988,Shih1988,Tapster1994,Rowe2001,Kocsis2011,Shadbolt2014}.

The problem of introducing a physically unambiguous separation of the total angular momentum (TAM)
of photons into a spin part (SAM) and an orbital part (OAM) is a controversial and debated subject~\cite{Akhiezer1965,VanEnk1994,Barnett1994,
Barnett2010,Bliokh2010,Bliokh2012,Franke2008}
since it was first proposed by J. Humblet in 1943~\cite{Humblet1943}.

The root of this long--standing problem lies in the transversality condition on the electromagnetic
field and the single--photon wavefunction, that introduces an interdependence between the spatial
and internal degrees of freedom of those quantities, hampering the possibility to define spin and
orbital rotations separately.


It is well known from both the quantum mechanical first--quantization description
of the photon \cite{Akhiezer1965,Berestetskii1982} and the classical electromagnetic theory
\cite{Soper1976,Jackson1998,VanEnk1994} that only the total angular momentum $\hat{\vett{J}}$ of
light is a well--defined and physically relevant quantity, subject to a conservation law
stemming from rotational invariance.
$\hat{\vett{J}}$ can be separated in two parts $\hat{\vett{L}}$ and $\hat{\vett{S}}$ that satisfy the
commutation relations characterizing the Lie algebra $so(3)$, thus representing
the correct generators of orbital and spin rotations, respectively.
However, $\hat{\vett{L}}$ and $\hat{\vett{S}}$ are both inconsistent with the transversality condition,
i.e. they do not leave the subspaces of longitudinal and transversal wavefunctions invariant, this fact
leads to difficulties in their physical interpretation.


The problem of providing an alternative to the canonical decomposition $\hat{\vett{J}} = \hat{\vett{L}}
+ \hat{\vett{S}}$ of the TAM was addressed in the
second--quantization framework and in the paraxial limit by Van Enk and Nienhuis~\cite{VanEnk1994}. Bliokh {\em{et al.}} in Ref.~\cite{Bliokh2010,Bliokh2012} continued this discussion within the first--quantization framework and beyond the paraxial limit.
In both cases, the authors proposed an alternative or non-canonical decomposition
$\hat{\vett{J}} = \hat{\vett{L}}' + \hat{\vett{S}}'$ of the TAM, in which the new orbital and spin
components are consistent with the transversality condition~\cite{Akhiezer1965,Barnett1994} and
therefore directly measurable. However, $\hat{\vett{L}}'$ and $\hat{\vett{S}}'$ do not satisfy the
commutation relations of the $so(3)$ algebra, therefore no longer representing the generators of rotations
in spatial and internal degrees of freedom, respectively.

In a recent work~\cite{GML2015JphysA} we presented a general formalism based on Kraus operators~\cite{Kraus1977} which allows in treating every
single-photon observable, including position and spin or momentum and helicity, in a unified picture. Consequently, making it possible to
construct the corresponding probability distributions in terms of Positive Operator-Valued Measures
(POVMs), a tool of paramount importance in the fields of quantum information science and open quantum
systems. In particular, we show how the transversality condition categorizes single--photon
observables into two classes: observables that are consistent with the transversality condition (e.g.
momentum, energy and helicity) are {\em{sharp quantum observables}}, described by Projector-Valued
Measures (PVMs), while observables that are not consistent with the transversality condition find
a natural description in terms of POVMs, i.e. they are {\em{unsharp quantum observables}}~\cite{Lahti1995,Holevo1982}. Considering the Heisenberg product $\Delta X$ $\Delta P$ this explains the increase via the unsharpness of the position observable.




The purpose of the present work is instead to face the problem of separating the TAM into a spin and
an orbital component, and of consistently describing these quantities in terms of PVMs and POVMs, in a
unified picture with position, spin, momentum and helicity.
We show that our generalization of Kraus' treatment allows to treat both the canonical and the
non--canonical decomposition of the TAM in a consistent way, endowing them with a clear quantum
information--theoretical characterization.

%
%

In particular, we find that the canonical OAM and SAM, $\hat{\vett{J}} =\hat{\vett{L}}+\hat{\vett{S}}$, are {\em{mutually compatible but unsharp}}
quantum observables, and we provide the explicit expression for the POVM describing their joint
measurements.
On the other hand, the non-canonical OAM and SAM, $\hat{\vett{J}} =\hat{\vett{L'}}+\hat{\vett{S'}}$, represent
{\em{mutually incompatible but sharp}} quantum observables, reflecting their consistency with the
transversality condition for single--photon wavefunctions.
Finally, we give explicit examples for both decompositions.

The paper is organized as follows.
In Section~\ref{sec:Formalism} the properties of the single--photon Hilbert space and the definition
of single--photon observables as POVMs are briefly recalled.
In Section~\ref{sec:photang} the TAM observable is presented, and its canonical and non--canonical
decomposition in an OAM and in a SAM part are discussed in detail.
The differences between the two decompositions are assessed with a study of Gaussian states in Section~\ref{sec:Results}, and Conclusions are drawn in the last Section \ref{sec:conclusions}.

\setlength{\arrayrulewidth}{0.5pt}
\begin{table}
\begin{center}
\begin{tabular}{||c|c|c||}
\hhline{|t===t|}
\multicolumn{3}{||c||}{\large{Total Angular Momentum (TAM)=SAM+OAM}}\\
\hhline{||=|=|=||}
&\cellcolor{blue!50}canonical&\cellcolor{green!50} non-canonical\\
\hhline{||-|-|-||}
& $\hat{J}=\hat{S}+\hat{L}$&$\hat{J}=\hat{S}'+\hat{L}'$\\
\hhline{||-|-|-||}
Generators of spatial and internal rotations& Yes$\quad\longrightarrow\quad[\hat{S}_k,\hat{L}_k]=0$&No$\quad\longrightarrow\quad[\hat{S}'_k,\hat{L}'_k]\not=0$\\
\hhline{||-|-|-||}
Transversality condition satisfied&No$\quad\longrightarrow\quad$ POVMs (unsharp observables)&Yes$\quad\longrightarrow\quad$ PVMs (sharp observables)\\
\hhline{|b:===:b|}
\end{tabular}
\end{center}
\caption{This table summarizes the main properties of the two decompositions of the sharp total angular momentum of a single photon.}
\label{tab:TAM}
\end{table}

\section{Single-photon states and observables}
\label{sec:Formalism}

In the present Section we briefly recall the description of single-photon observables in terms of POVMs, which generalizes the treatment of the position observable proposed by K. Kraus \cite{Kraus1977}. The interested reader is referred to \cite{GML2015JphysA} for all the details in merit.

The proper quantum-mechanical description of a single photon in free space takes place in the Hilbert space 
\begin{eqnarray}
\label{thespace}
\hsys 
&=&
\left\{ \bgreek{\psi}(\vett{p}) : \bgreek{\psi}(\vett{p}) =
\sum_{i=1}^2 \psi^i(\vett{p}) \otimes \vett{e}_i(\vett{p}) \, \biggl{|} \, \vett{e}_i(\vett{p})\in\mathbb{C}^3\;\textrm{and}\;
\psi^i(\vett{p}) \in
\mathcal{L}^2\left( \R^3,\frac{d^3\vett{p}}{|\vett{p}|}\right)
\right\},
\end{eqnarray}
where the set $\lbrace \vett{e}_{i}(\vett{p})\rbrace_{i=1,2,3}$ denotes
the so-called \emph{intrinsic frame}, i.e. a reference frame such that one of the axis is directed along the direction of the momentum, i.e.
$\vett{e}_3(\vett{p}) = \frac{\vett{p}}{|\vett{p}|}$, and
$\vett{e}_1(\vett{p}) \times \vett{e}_2(\vett{p}) = \vett{e}_3(\vett{p})$.
$\hsys$ as defined in Eq. \eqref{thespace} is equipped with the positive definite inner product
\begin{equation}\label{theinnerproduct}
\braket{ {\bgreek{\phi}} |  \bgreek{\psi} }:=
\int \frac{d^3\vett{p}}{|\vett{p}|} \, \left(
\phi^{1}(\vett{p})^* \, \psi^1(\vett{p}) +
\phi^{2}(\vett{p})^* \, \psi^2(\vett{p}) \right)\;.
\end{equation}
Note that $\hsys$ defined in Eq. \eqref{thespace} is isomorphic to $\mathcal{L}^2\left( \R^3,\frac{d^3\vett{p}}{|\vett{p}|}\right) \otimes \mathbb{C}^2 $ despite the photon spin number is $s=1$. This traduces the well-known transversality condition, i.e. the condition $ \bgreek{\psi}(\vett{p}) \cdot \vett{p} = 0$ according to which the longitudinal component of the photon state, namely the one along the direction of $\vett{e}_3(\vett{p})$, is suppressed~\cite{Wigner1939,Moses1967,Moses1968,Kraus1977,Kraus1977,
GML2015JphysA}. Consequently, the degrees of momentum and spin get entangled in a fuddling way, which leads to the rich physics of photons.

We stress that this result can be derived by only requiring the
single-photon Hilbert space to carry an irreducible representation of
the Poincar\'e group uniquely characterized by spin $s=1$ and mass $m=0$
Casimir invariants. In particular, the mass-shell condition, which also implies the transversality condition, naturally selects $\hsys$ as the proper subspace of a spin $s=1$ irreducible representation of Poincar\'e group carrying a positive definite inner product (the latter being a necessary ingredient in order to endow the theory of a probabilistic character) \cite{GML2015JphysA}.

The main idea behind a neat treatment of all single photon observables in a unified way is to formalize the suppression of the longitundinal component of the wavefunction in terms of the action of a projection operator $\hat{\YUPPI}(\vett{p})  : \ha \to \hsys$
\begin{align}\label{GB2}
&\YUPPI^j_{\,\,k}(\vett{p}) = \delta^j_{\,\,k} - \frac{p^j p_k}{|\vett{p}|^2} \qquad \forall \, \vett{p} \in \R^3\;,
\end{align}
where
\begin{equation}
\label{theextendedspace}
\ha 
 = \left\{ \mathbf{f}(\vett{p}) : \mathbf{f}(\vett{p}) =
\sum_{i=1}^3 \psi^i(\vett{p}) \otimes \vett{e}_i(\vett{p}) \, \biggl{|} \, \vett{e}_i(\vett{p})\in\mathbb{C}^3\;\textrm{and}\;\,
\psi^i(\vett{p}) \in \mathcal{L}^2\left( \R^3,\frac{d^3\vett{p}}{|\vett{p}|}\right)
\right\},
\end{equation}
is isomorphic to $\simeq \mathcal{L}^2\left( \R^3,\frac{d^3\vett{p}}{|\vett{p}|}\right) \otimes \mathbb{C}^3$ and consists of wave-functions that differ from those of $\hsys$ simply by the presence of a longitudinal component.
Physically, the projector \eqref{GB2} can be interpreted as a quantum analogue of the Helmholtz projection which is used to decompose the electric and magnetic field into a longitudinal and a transversal component.

The introduction of the Hilbert space $\mathscr{H}_A $ is the key for a unified treatment of all single-photon observables.
In particular, any observable $\mathcal{O}$ can be associated to a self-adjoint operator $\hat{O} $ defined upon it which remarkably retains the same structure as in the case of a relativistic massive spin $s=1$ particles \cite{Wigner1939,Moses1967,Moses1968,Kraus1977,Kraus1977}, opportunely adapted to the massless case of photons by the constraint $p^0 = |\vett{p}|$.

This means, for example, that
\begin{equation}
\label{gamma_mom}
\begin{split}
\hat{P}_k      \, \mathbf{f} (\vett{p}) &=
p_k \, \mathbf{f}(\vett{p}) \\ 
\hat{X}^{NW}_k \, \mathbf{f} (\vett{p}) &=
      i \hbar \, \frac{\partial  \mathbf{f}(\vett{p})}{\partial p_k} +
\frac{i \hbar}{2} \, \frac{p_k}{|\vett{p}|^2} \, \mathbf{f}(\vett{p}) \\ 
\hat{J}_k      \, \mathbf{f} (\vett{p}) &= S_k \vett{f}(\vett{p}) +
\left( i \hbar \, \partial_\vett{p} \times \vett{p} \right)_k \vett{f}(\vett{p}) , 
\end{split}
\end{equation}
represent the momentum (generator of spatial translations), Newton-Wigner position (generator of boosts) and TAM (generator of rotations) operators, respectively.

In general, given that the system is described by a state $\ket{\phi}$, the probability for any observable $\mathcal{O}$ to have an outcome in a generic measurable set $\mathcal{M}$ belonging to a suitable $\sigma-$algebra is given by the familiar expression
\begin{equation}
p\left( \mathcal{O} \in \mathcal{M} \right) = \bra{\phi}\hat{E}_{\mathcal{O}}(\mathcal{M}) \ket{\phi},
\end{equation}
where $\hat{E}_{\mathcal{O}}(\mathcal{M})$ is the associated PVM.
When we consider photons, we need to move from the extended Hilbert space $\mathscr{H}_A $, where all these observables are well-defined, to the physical Hilbert space $\hsys$ in order to cope with the transversality condition \eqref{thespace}. Thus projecting the associated PVM $\mathcal{M} \mapsto \hat{E}_{\mathcal{O}}(\mathcal{M})$ defined on $\ha$ onto $\hsys$ through the operator $\hat{\YUPPI}$, we obtain
\begin{equation}\label{probdistribPOVM}
p\left(\mathcal{O}\in\mathcal{M}\right) = \bra{\psi}\hat{F}_{\mathcal{O}}(\mathcal{M})\ket{\psi},
\end{equation}
where $\ket{\psi} \in \hsys$ describes the single-photon wave-function and
\begin{equation}
\label{pvm2povm}
\hat{F}_{\mathcal{O}}(\mathcal{M})\; =\;
\hat{\YUPPI} \hat{E}_{\mathcal{O}}(\mathcal{M})
\hat{\YUPPI}  \;=\; \hat{\Omega}^{\dagger}_{\mathcal{O}}(\mathcal{M}) \hat{\Omega}_{\mathcal{O}}(\mathcal{M}),
\qquad\textrm{with}\quad
\hat{\Omega}_{\mathcal{O}}(\mathcal{M}) = \hat{E}_{\mathcal{O}}(\mathcal{M}) \hat{\YUPPI}\;.
\end{equation}
The resulting map $\mathcal{M} \mapsto \hat{F}_O(\mathcal{M})$ is a \emph{Positive Operator-Valued Measure} (POVM), a
well-known concept and a widely-used tool in quantum information and open quantum systems theory.
 Let us remind the reader that the variance of a POVM in contrast to a PVM loses the idempotence property, i.e. $\hat{F}^2_O(\mathcal{M}) \neq \hat{F}_O(\mathcal{M})$.
Observables described by POVMs are referred to as \emph{unsharp}~\cite{Lahti1995,Busch1996,Busch2009,Busch2010}, since their emergence reflects either practical limits in the precision of measurements (in which case POVMs are coarse--grained versions of PVMs)~\cite{Hawton2010,Mandel1995} or the inherent impossibility of realizing a preparation in which the value
of an observable is perfectly defined~\cite{Holevo1982,Lahti1995,Busch1996}.
This statement can be made more quantitative by simply showing that
\begin{equation}
\label{eq:increase}
\begin{split}
\mbox{Var}_{\mbox{povm}}(\mathcal{O}) &= \,
  \bra{\psi} \hat{\YUPPI} \, \hat{O}^2  \hat{\YUPPI} \ket{\psi}
- \bra{\psi} \hat{\YUPPI} \,  \hat{O}  \hat{\YUPPI} \ket{\psi}^2 \\
&= \,
  \mbox{Var}_{\mbox{pvm}}(\mathcal{O})
+ \bra{\psi} \left( \hat{\YUPPI} \, \hat{O}^2 \hat{\YUPPI}
             -    ( \hat{\YUPPI} \, \hat{O} \hat{\YUPPI} )^2 \right) \ket{\psi}  \\
&= \,
  \mbox{Var}_{\mbox{pvm}}(\mathcal{O}) + \bra{\phi} \left( 1 - \hat{\YUPPI} \right) \ket{\phi}
\, ,
\end{split}
\end{equation}
where $\ket{\phi} = \hat{O} \, \hat{\YUPPI}|\psi\rangle$.
Since the operator $1 -\hat{\YUPPI}$ is positive, the variance
$\mbox{Var}_{\mbox{povm}}(\mathcal{O})$ is always larger than the variance
$\mbox{Var}_{\mbox{pvm}}(\mathcal{O})$ that would arise if the POVM
operators were idempotent, i.e. in the case of a PVM.
In this sense, POVMs increase the statistical character of quantum observables \cite{Massar2007}.
A final remark is worth to be made at this point concerning Eq. \eqref{eq:increase}.
It is evident in fact that if
\begin{equation}
\label{pvm2povm_caveat}
\left[ \hat{O} , \hat{\YUPPI}  \right] = 0,
\end{equation}
the second term on its r.h.s. vanishes. In this case we say that the observable $\mathcal{O}$ is \textit{compatible} with the transversality condition.
Examples of observables which are compatible with the transversality condition (and thus sharp) are momentum and helicity, while examples of observables which are incompatible with Eq. \eqref{pvm2povm_caveat} (and thus unsharp) are position and spin \cite{GML2015JphysA}.

The incompatibility with the transversality condition has led to the introduction of opportunely modified position \cite{Hawton1999} and spin \cite{VanEnk1994, Bliokh2010} operators. It is quite straightforward to show that such modified operators correspond to the projected version, through $\hat{\YUPPI}$, of the familiar operators defined in Eq. \eqref{gamma_mom}.
Such modified operators thus become compatible \textit{by construction} with Eq. \eqref{pvm2povm_caveat} but no longer represent the generators of boosts and rotations, respectively.

Moreover, if $\hat{F}_{O}(\mathcal{M})$ is idempotent, then
\begin{equation}
\left[ \hat{E}_{O}(\mathcal{M}) , \hat{\YUPPI}  \right] = 0
\end{equation}
for all $\mathcal{M}$ and thus Eq.\eqref{pvm2povm_caveat} holds.
Indeed, one can write
\begin{eqnarray}
\hat{F}_{O}(\mathcal{M}) &=& \hat{\YUPPI} \hat{E}_{O}(\mathcal{M}) \hat{\YUPPI} + \hat{\YUPPI}  \hat{E}_{O}(\mathcal{M}) \hat{\YUPPI}^\perp + \hat{\YUPPI}^\perp \hat{E}_{O}(\mathcal{M}) \hat{\YUPPI} + \hat{\YUPPI}^\perp \hat{E}_{O}(\mathcal{M}) \hat{\YUPPI}^\perp ,
\end{eqnarray}
with $\hat{\YUPPI}^\perp= 1 -  \hat{\YUPPI}$. The idempotence of $\hat{F}_{O}(\mathcal{M})$ implies
$\hat{\YUPPI}  \hat{E}_{O}(\mathcal{M}) \hat{\YUPPI}^\perp = 0$, and thus $\left[ \hat{E}_{O}(\mathcal{M}) , \hat{\YUPPI}  \right] = 0$.

It is finally worth remarking that there may exist specific states $\ket{\psi^*} \in \hsys$ such that the mean value $ \bra{\psi^*} \left[ \hat{O} , \hat{\YUPPI}  \right] \ket{\psi^*} = 0$, even though Eq.\eqref{pvm2povm_caveat} is not satisfied. This is a general property of the uncertainty relations of the Robertson form~\cite{Robertson} and can be circumvent by transforming uncertainty relations to entropic ones~\cite{entropicuncertainty,entropicuncertainty2}. If a single photon is prepared in such state, the extra variance in \eqref{eq:increase} disappears despite $\mathcal{O}$ being unsharp.

In the next Sections \ref{sec:photang}, \ref{sec:photang2} we will show how this final subtle point can be related to the behavior of the canonical and non-canonical decompositions of the TAM in the paraxial limit. We will now see how this formalism, when applied to single photon angular momentum, provides the two decompositions with a novel and unified interpretation in a quantum-information perspective and allows to evaluate the corresponding probability distributions according to Eq.~\eqref{pvm2povm}.

%
%

\section{Canonical decomposition of the total angular momentum}
\label{sec:photang}

Since the main focus of the present work is on the angular momentum, it is essential to remind that $\hsys$ hosts the irreducible representation of the roto-translation group
\begin{equation}
\label{eq:rototr_hS}
\hat{U}(\vett{a},R)\;  \bgreek{\psi}(\vett{p})\; =\;
e^{- \frac{i}{\hbar} \vett{a} \cdot \vett{p} } \,
R \;  \bgreek{\psi}(R^{-1} \vett{p}),
\end{equation}
with $R \in SO(3)$ being a rotation matrix and $\vett{a} \in \mathbb{R}^3$
a translation vector.
This representation is consistently maintained also on the extended Hilbert space $\ha$, where all the spin and angular momentum operators can be properly defined.

The spin $s=1$ of the photon has a deep consequence on the connection between spin and rotations (and thus between internal and configurational degrees of freedom), which is compressed in the following key relation
\begin{equation}
\label{su2so3}
R = \hat{V}^{\dagger}\, e^{-\frac{i}{\hbar}\, \varphi\, \vett{n} \cdot \vett{\hat{S}}}\, \hat{V},
\end{equation}
where $\hat{\vett{S}}$ are the generators of the $SO(3)$ vector rotations~\cite{NoriReview} and $\hat{V}$ is an opportune unitary matrix.
The matrix $\hat{V}$ has also the remarkable role to show the equivalence between
the condition of transversality~\eqref{thespace} and that of non-zero
helicity, which is also known to characterize single-photon wavefunctions~\cite{GML2015JphysA}.
In fact a straightforward calculation shows that the eigenfunctions of the
helicity operator
\begin{equation}
\label{unirep1}
\hat{\epsilon} =
\frac{1}{\hbar} \, \frac{\vett{\hat{S}} \cdot \vett{p}}{|\vett{p}|},
\end{equation}
are given by $\hat{V}\vett{e}_3(\vett{p})$ with eigenvalue $0$ and $\hat{V} \tilde{\bf{e}}_\pm(\vett{p}) = \hat{V}\, \frac{ \vett{e}_1(\vett{p}) \mp i
\vett{e}_2(\vett{p}) }{\sqrt{2}}$ with eigenvalue $\pm1$.
The suppression of the longitudinal component is therefore unitarily equivalent to the suppression of zero-helicity eigenstates.

It is important now to notice that any choice of a particular representation of these generators such that the $su(2)$ algebra is satisfied (i.e. $\left[\hat{S}_i, \hat{S}_j\right] = \epsilon_{ijk}\, \hat{S}_k$) uniquely determines the matrix $\hat{V}$ according to \eqref{su2so3}.
Equivalently said, \eqref{su2so3} uniquely fixes the couple $\left( \vett{\hat{S}}, \hat{V}\right)$.
As an example, if we choose the spin matrices to be of the form
\begin{equation}
\label{eq:paulimat}
\hat{S}_x = \frac{\hbar}{\sqrt{2}}
\begin{pmatrix}
0 & 1 & 0 \\
1 & 0 & 1 \\
0 & 1 & 0
\end{pmatrix} \quad\quad
\hat{S}_y = \frac{\hbar}{\sqrt{2}}\begin{pmatrix}
 0 & -i &  0 \\
 i &  0 & -i \\
 0 &  i &  0
\end{pmatrix} \quad\quad
\hat{S}_z = \,\hbar\,\begin{pmatrix}
1 & 0 & 0 \\
0 & 0 & 0 \\
0 & 0 & -1
\end{pmatrix}
\end{equation}
then $\hat{V}$ is equal to
\begin{equation}
V = \begin{pmatrix}
\frac{1}{\sqrt{2}} & -\frac{i}{\sqrt{2}} & 0 \\
0 & 0 & -1 \\
-\frac{1}{\sqrt{2}} & -\frac{i}{\sqrt{2}} & 0 \\
\end{pmatrix}.
\end{equation}
Alternatively, we can choose $\hat{V}= \Id$, this way fixing the three relevant spin matrices to have the following representation
\begin{equation}\label{OurSpinMatricesNow}
\hat{S}_x =\,\hbar\\,
\begin{pmatrix}
0 & 0 & 0 \\
0 & 0 & -i \\
0 & i & 0
\end{pmatrix} \quad\quad
\hat{S}_y =\,\hbar\\,\begin{pmatrix}
 0 & 0 &  i \\
 0 &  0 & 0 \\
 -i &  0 &  0
\end{pmatrix} \quad\quad
\hat{S}_z = \,\hbar\,\begin{pmatrix}
0 & -i & 0 \\
i & 0 & 0 \\
0 & 0 & 0
\end{pmatrix}\;,
\end{equation}
which has the merit to immediately show that they represent the anti-symmetric subset of the $su(3)$ Gell-Mann matrices.
This obviously has to be the case, because only in this case $e^{\frac{i}{\hbar} \phi\,\vec{n}\cdot\mathbf{\hat{S}}}$ becomes real and, consequently, can be identified with the rotation matrix $R$ in the real space.

As discussed in the previous Section, the generator of rotations, i.e. the single--photon TAM,
can be canonically decomposed on $\ha$ via
\begin{equation}\label{TAM}
\hat{J}_k\;\vett{f}(\vett{p}) = (\mathbf{\hat{S}}+\mathbf{\hat{L}})_k \;\vett{f}(\vett{p})\;=\;\hat{S}_k\; \vett{f}(\vett{p}) + \left( i \hbar \, \partial_\vett{p} \times \vett{p} \right)_k\; \vett{f}(\vett{p})
\quad\quad\quad k=1,2,3\\.
\end{equation}
The OAM, $\hat{\mathbf{L}}$, and SAM, $\hat{\mathbf{S}}$, involved in the canonical decomposition obey the familiar angular momentum commutation
relations
\begin{equation}
\label{eq:intro1}
[ \hat{S}_i , \hat{S}_j ] = i\hbar \, \sum_{k=1}^3 \epsilon_{ijk} \, \hat{S}_k
\quad , \quad
[ \hat{L}_i , \hat{L}_j ] = i\hbar \, \sum_{k=1}^3 \epsilon_{ijk} \, \hat{L}_k
\quad , \quad
[ \hat{S}_i ,\hat{L}_j ] = 0
\;,
\end{equation}
characterizing the Lie algebra $so(3)$ of the rotation group $SO(3)$, thus allowing to regard them as the generators of  internal and spatial rotations, respectively.

In the light of~\eqref{gamma_mom}, we can also consistently express the canonical OAM in terms of the Newton-Wigner
position operator as
\begin{equation}
\hat{\vett{L}} = \hat{\vett{X}}_{NW} \times \hat{\vett{P}}\;.
\end{equation}
This form makes immediately evident that the OAM and SAM involved in the canonical decomposition
of the TAM satisfy equation~\eqref{pvm2povm_caveat} due to the presence of the Newton-Wigner
position operator and of the Pauli matrices, respectively. The calculation of the commutators $[\hat{S}_k,\hat{V}\YUPPI \hat{V}^\dag]$, $[\hat{L}_k,\hat{V} \YUPPI \hat{V}^\dag]$, explicitly showing the intrinsic unsharpness of both of them, is given in Appendix~\ref{app:commutators}.

The canonical OAM and SAM are therefore inconsistent with the transversality condition,
except in the paraxial limit, where the wavefunction $\bgreek{\psi}(\vett{p})$ is concentrated around a certain value $\vett{p}_0$. To make this statement quantitative, let us consider a family of wavefunctions $\bgreek{\psi}_a(\vett{p}) = \vett{u}
(\vett{p}) \, f_a(\vett{p}) \in \hsys$ such that $\vett{p} \cdot \vett{u}(\vett{p}) = 0$ (to comply with the transversality condition), normalized $\| \vett{u}(\vett{p}) \|^2 = 1$ and which satisfies $\lim_{a \to 0}
|f^2_a(\vett{p})|^2 = |\vett{p}_0| \, \delta(\vett{p}-\vett{p}_0)$.
For the observable $\hat{\mathbf{S}} \cdot \vett{n}$, the extra variance corresponding to the second term of~\eqref{pvm2povm_caveat} reads
\begin{equation}
\begin{split}
&\bra{\phi_a} 1 - \hat{V}^\dag \YUPPI \hat{V} \ket{\phi_a} =
\int \frac{d^3\vett{p}}{|\vett{p}|} \, |f^2_a(\vett{p})|^2 \,
\Big( \vett{\hat{S}} \cdot \vett{n} \, \hat{V} \YUPPI(\vett{p}) \, \vett{u}(\vett{p}) \Big)^*
\Big( 1-\hat{V}^\dag \YUPPI(\vett{p}) \hat{V} \Big)
\Big( \vett{\hat{S}} \cdot \vett{n} \, \hat{V} \YUPPI(\vett{p}) \, \vett{u}(\vett{p}) \Big) \\
\\
\lim_{a \to 0} \,
&\bra{\phi_a} 1 - \hat{V}^\dag \YUPPI \hat{V} \ket{\phi_a} =
\Big( \vett{\hat{S}} \cdot \vett{n} \, \hat{V} \, \vett{u}(\vett{p}_0) \Big)^*
\Big(1-\hat{V}^\dag\YUPPI(\vett{p}_0)\hat{V} \Big)
\Big( \vett{\hat{S}} \cdot \vett{n} \, \hat{V} \, \vett{u}(\vett{p}_0) \Big) \\
\end{split}
\end{equation}
so that, if $\hat{V} \vett{u}(\vett{p}_0)$ is an eigenfunction of $\vett{\hat{S}} \cdot \vett{n}$ with eigenvalue $\pm 1$, \begin{equation}\lim_{a \to 0} \bra{\phi_a} 1 - \hat{V}^\dag \YUPPI \hat{V} \ket{\phi_a} = 0\;.\end{equation}
An identical observation characterizes the canonical OAM in the paraxial limit.

Finally, let us remark that the TAM defined in \eqref{TAM}, despite being given by the sum of two unsharp observables,
is a sharp observable consistent with the transversality condition \eqref{pvm2povm_caveat} (the proof of this fact can be found in
Appendix~\ref{app:commutators}). Therefore, its statistics is described in terms of a PVM. An example is provided in section~\ref{sec:JointSzLzCan}.

%

\section{Non-canonical decomposition of the TAM}
\label{sec:photang2}

As stressed before, the operators $\hat{L}_k$ and $\hat{S}_k$ which stem from the canonical decomposition of the TAM are not compatible with the transversality (or equivalently with the non-zero helicity) condition, which means that their action on a transverse wavefunction (i.e. a physical singe photon state) results in a non-vanishing longitudinal component \cite{Akhiezer1965,VanEnk1994,Bliokh2010}.
This fact has led to introduce an alternative decomposition of the TAM in such a way that the two resulting components would be consistent with the transversality condition, therefore becoming direct observables on the physical single-photon Hilbert space $\hsys$, but no longer representing the generators of rotations and translations~\cite{Berry2009,Li2009,Bliokh2010}.

We recall here this non-canonical decomposition and endow it with a clear interpretation in a quantum-information theoretical perspective.
Making use of the identity
\begin{equation}
\left( \vett{p} \times ( \vett{p} \times \hat{\vett{S}}) \right)_k = p_k \, (\vett{p} \cdot \hat{\vett{S}}) - |\vett{p}|^2 \, \hat{S}_k,
\end{equation}
we have that $\hat{J}_k = \hat{L}^\prime_k + \hat{S}^\prime_k$, where
\begin{equation}
\label{eq:dec_alternatA}
\hat{S}^\prime_k = \frac{p_k }{|\vett{p}|} \, \left( \frac{\vett{p}}{|\vett{p}|} \cdot \hat{\vett{S}} \right) = \hbar \,\frac{p_k }{|\vett{p}|} \, \hat{\epsilon}
\end{equation}
and
\begin{equation}
\label{eq:dec_alternatB}
\hat{L}^\prime_k = \hat{L}_k  - \frac{\left( \vett{p} \times (\vett{p} \times \hat{\vett{S}}) \right)_k}{|\vett{p}|^2}.
\end{equation}
In equation~\eqref{eq:dec_alternatA} $\hat{\epsilon}$ denotes the helicity operator defined in equation~\eqref{unirep1}. We stress that both observables are well-defined in terms of self-adjoint operators only on the extended Hilbert space $\hbarra_A$.
Since the operator $\hat{\mathbf{S}'}$ satisfies the transversality condition~\eqref{pvm2povm_caveat}, also $\hat{\mathbf{L}'}$ has to have a vanishing commutation relation with the projection onto the physical Hilbert space, i.e.
\begin{equation}
\left[ \hat{V}^\dag \hat{L}'_k \hat{V} , \hat{\YUPPI} \right] =
\left[ \hat{V}^\dag \hat{J}_k \hat{V} , \hat{\YUPPI} \right] -
\left[ \hat{V}^\dag \hat{S}'_k \hat{V} , \hat{\YUPPI} \right] =
0\;.
\end{equation}
Both $\hat{S}'_k$ and $\hat{L}'_k$ therefore classify as \emph{sharp} observables~\cite{Lahti1995,Busch2010} and their probability distributions are simply obtained as the mean values of the associated family of PVMs.

The non-canonical decompositon in equations~\eqref{eq:dec_alternatA}--\eqref{eq:dec_alternatB} has several remarkable differences with respect to the canonical one. First of all, the components of the SAM operator $\hat{S}'_k$ commute with each other, while the components of the OAM operator $\hat{L}'_k$ do not commute with each other. Thus the rotation and translation degrees of freedoms are no longer treated on equal footing.
Moreover, the SAM and OAM operators $\hat{S}'_k$ and $\hat{L}'_k$ do not commute with each other, thus being mutually incompatible quantum observables~\cite{Bliokh2010},  in strong contrast to the canonical decomposition into SAM and OAM.
The explicit expressions of the commutators $[\hat{L}'_i,\hat{S}'_j]$, $[\hat{L}'_i,\hat{L}'_j]$ are given in Appendix~\ref{app:commutators}.

It is finally important to point out that, despite the decomposition of the TAM into a SAM and an OAM part is highly not unique and further different decompositions could be taken into account, we have analyzed the two most relevant from the physical point of view. The canonical decomposition of the TAM is in fact dictated by the additional constraint that the resulting OAM and SAM represent the correct generators of spatial and internal rotations. The non-canonical decomposition is instead fixed by the constraint that both components are compatible with the transversality condition,  i.e. for Eq.~\eqref{pvm2povm_caveat} to be satisfied, which for the SAM part means that the eigenspace relative to the eigenvalue $0$ coincides with that of helicity.


\section{Implications onto observables}

We illustrate the two different decompositions now by two explicit physical examples, i.e. for photons with a fixed spatial direction and for Gaussian states with definite circular polarization.

\subsection{Joint POVM of the canonical SAM and OAM along a fixed spatial direction}
\label{sec:JointSzLzCan}

We have shown that both the SAM and OAM are unsharp observables for which a PVM consistent with the transversality
condition cannot be given (see also equation~\eqref{eq:commutators_AM} of Appendix~\ref{app:commutators} for details).
In the present Section we explicitly derive the joint probability distribution and marginals of the single--photon
OAM and SAM along a generic spatial axis.
This is achieved through the concrete construction of the join POVMs of such observables, according to the procedure
outlined in \cite{GML2015JphysA} and in the previous Section.
In the following calculation we consider, without any loss of generality, the spin and orbital angular momenta along
the $z$-axis of a suitable Cartesian frame in real space.
We make use of spherical coordinates $(p,\theta,\phi)$ relative to the $z$-axis in momentum space, denoting through
$\{ \vett{e}_s \}_{s=1}^3$ the canonical basis of $\mathbb{R}^3$.

On the extended Hilbert space $\ha$, the joint eigenfunctions of the compatible observables $\hat{L}_z$ and $\hat{S}_z$ have the familiar form
\begin{equation}
\label{eq:eigenLS}
\vett{u}_{m,m_s}(\vett{p}) = f(p, \theta) \, \frac{e^{im\phi}}{\sqrt{2\pi}} \, \vett{e}_{s},\quad 
\hat{L}_z \vett{u}_{m,m_s} = \hbar m \vett{u}_{m,m_s}, \quad \hat{S}_z \vett{u}_{m,m_s} = \hbar m_s \vett{u}_{m,m_s} ,
\end{equation}
where $m_s = 2-s$, and $f(p, \theta)$ is a square-integrable function of the variables $p,\theta$ ensuring the proper normalization of \eqref{eq:eigenLS}
\begin{equation}
\int_0^{\infty} dp \,\, p \, \int_0^\pi d\theta \, \cos(\theta) \, |f(p, \theta)|^2 = 1\; .
\end{equation}
The joint PVM of $\hat{L}_z$ and $\hat{S}_z$ is then given by
\begin{equation}
\label{eq:pvmLS}
(m,m_s) \mapsto
\left( \hat{E}_{L_z,S_z}(m,m_s) \, \vett{f} \right)(\vett{p}) =
\frac{e^{im\phi}}{\sqrt{2\pi}} \, \vett{e}_{s}
\,\,
\int_0^{2\pi} d\phi' \,
\frac{e^{- im\phi'}}{\sqrt{2\pi}} \, \vett{e}_{s}^* \cdot \vett{f}(p,\theta,\phi')
\quad .
\end{equation}
The correspondent POVM on the physical single-photon Hilbert space $\hsys$ is therefore obtained by the
application of Eq.~\eqref{pvm2povm} and reads
\begin{equation}
\label{eq:povmLS}
(m,m_s) \mapsto
\left( \hat{F}_{L_z S_z}(m,m_s) \, \bgreek{\psi} \right)(\vett{p}) =
\frac{e^{im\phi}}{\sqrt{2\pi}} \,
\YUPPI({\vett{p}}) \hat{V}^\dag \vett{e}_{s}
\,
\left(
\int_0^{2\pi} d\phi' \,
\frac{e^{- im\phi'}}{\sqrt{2\pi}} \, \vett{e}_{s}^* \cdot \hat{V} \bgreek{\psi}(p,\theta,\phi') \right)
\quad .
\end{equation}
The joint probability distribution of such observables is readily obtained using Eq.~ \eqref{probdistribPOVM}
\begin{equation}
\label{eq:ddpLS}
p_{L_z,S_z}(m,m_s) = \int_0^\infty dp \, p \, \int_0^{\pi} d\theta \, \sin{\theta} \,
\left|
\int_0^{2\pi} d\phi' \,
\frac{e^{- im\phi'}}{\sqrt{2\pi}} \, \vett{e}_{s}^* \cdot \hat{V} \bgreek{\psi}(p,\theta,\phi')
\right|^2
\quad .
\end{equation}
The marginals of \eqref{eq:ddpLS} read
\begin{equation}
p_{L_z}(m) = \int_0^\infty dp \, p \, \int_0^{\pi} d\theta \, \sin{\theta} \,
\left|
\int_0^{2\pi} d\phi' \,
\frac{e^{- im\phi}}{\sqrt{2\pi}} \, \bgreek{\psi}(p,\theta,\phi')
\right|^2
\end{equation}
and
\begin{equation}
p_{S_z}(m_s) = \int_0^\infty dp \, p \, \int_0^{\pi} d\theta \, \sin{\theta} \,
\int_0^{2\pi} d'\phi \,
\left| \vett{e}_{s}^* \cdot \hat{V} \bgreek{\psi}(p,\theta,\phi') \right|^2
\quad .
\end{equation}

\subsection{PVMs of the non-canonical SAM and OAM along a fixed spatial direction}

As another example of the versatility and generality of the formalism outlined above, and for conformity with Section~\ref{sec:JointSzLzCan}, we explicitly derive the joint probability distribution of the non-canonical SAM and OAM operators along a fixed spatial axis, which again we consider to be the $z$-axis.
Let us begin by reminding that the vectors $\hat{V} \tilde{\vett{e}}_\pm(\vett{p})$, with $
\tilde{\vett{e}}_\pm(\vett{p}) = \frac{ \tilde{\vett{e}}_1(\vett{p}) \mp i \tilde{\vett{e}}_2(\vett{p}) }{\sqrt{2}} $, are the eigenvectors of the helicity operator $\hat{\epsilon}$ relative to the eigenvalues $\pm 1$, and depend on $\vett{p}$ only through the angles $\theta,\phi$.
Therefore, the joint generalized eigenfunctions of helicity and $ \hat{S}'_z$ are the following elements of $\hbarra_A$:
\begin{equation}
\label{eq:sam1}
\vett{u}_{\pm 1, s_0}(\vett{p}) = f(p,\phi) \, \delta(\cos(\theta)\mp s_0) \otimes \hat{V} \tilde{\vett{e}}_\pm(\vett{p})
\end{equation}
where $s_0 \in [-1,1]$ and $f(p,\phi)$ is a properly normalized function. It is clear from Eq.~ \eqref{eq:sam1} that
\begin{equation}
\hat{S}'_z \, \vett{u}_{\pm 1, s_0}(\vett{p}) = s_0 \, \vett{u}_{\pm \, s_0}(\vett{p})
\quad,\quad
\mathbf{\hat{\epsilon}}\; \vett{u}_{\pm 1, s_0}(\vett{p}) = (\pm 1) \, \vett{u}_{\pm 1, s_0}(\vett{p})\;.
\end{equation}
Notice that $\cos(\theta) = s_0$ and $\tilde{\vett{e}}_+(\vett{p})$ lead to $\hat{S}'_z = s_0$, as well as $\cos(\theta) =-s_0$ and
$\tilde{\vett{e}}_-(\vett{p})$.
As a consequence, the event $\hat{S}'_z \in \mathcal{M}$, where $\mathcal{M}$ is a Borel subset of $[-1,1]$, happens if and only if
$\cos(\theta) \in \mathcal{M}_+$ and $\epsilon = 1$ or
$\cos(\theta) \in \mathcal{M}_-$ and $\epsilon = -1$, the sets $\mathcal{M}_\pm$ being:
\begin{equation}
\mathcal{M}_\pm = \left\{ \theta : \pm \cos(\theta) \in \mathcal{M} \right\} \; .
\end{equation}
The probability that $\hat{S}'_z \in \mathcal{M}$ then clearly reads:
\begin{equation}
\label{eq:sam2}
p(\hat{S}'_z \in \mathcal{M}) = \int_0^\infty dp \, p \, \int_0^{2\pi} d\phi
\left( \int_{\mathcal{M}_+} d\theta \sin(\theta) \, | \hat{V} \tilde{\vett{e}}^*_+(\vett{p}) \cdot \vett{f}(\vett{p}) |^2 +
       \int_{\mathcal{M}_-} d\theta \sin(\theta) \, | \hat{V} \tilde{\vett{e}}^*_-(\vett{p}) \cdot \vett{f}(\vett{p}) |^2 \right)
\, ,
\end{equation}
where $\vett{f} \in \hbarra_A$. The probability can be written as the average
\begin{equation}
\label{eq:sam3}
p(\hat{S}'_z \in \mathcal{M}) = \langle \vett{f} | \hat{E}_{S'_z,\epsilon}(\mathcal{M},1) | \vett{f} \rangle +
                                \langle \vett{f} | \hat{E}_{S'_z,\epsilon}(\mathcal{M},-1) | \vett{f} \rangle
\end{equation}
over $\vett{f}$ of the projector $\hat{E}_{S'_z,\epsilon}(\mathcal{M},1) + \hat{E}_{S'_z,\epsilon}(\mathcal{M},-1) $.
In the light of \eqref{eq:sam3}, the joint PVM associated to $\hat{S}'_z$  and helicity on $\hbarra_A$ is
\begin{equation}
\label{eq:sam4}
(\mathcal{M},\pm 1) \mapsto \left( \hat{E}_{S'_z,\epsilon}(\mathcal{M},\pm 1) \, \vett{f} \right)(\vett{p}) = 1_{\mathcal{M}_\pm}(\theta) \,
\hat{V} \tilde{\vett{e}}_\pm(\vett{p})  \,
\Big( \hat{V} \tilde{\vett{e}}^*_\pm(\vett{p}) \cdot \vett{f}(\vett{p}) \Big)
\end{equation}
and the PVM associated to $\hat{S}'_z$ is the marginal of \eqref{eq:sam4} over the helicity degrees of freedom.

The projection \eqref{pvm2povm} of the PVM \eqref{eq:POVMlprime} onto the physical Hilbert space $\hsys$
\begin{equation}
\label{eq:POVMsprime}
\mathcal{M} \mapsto \left( \hat{F}_{S'_z, \epsilon}(\mathcal{M}) \, \bgreek{\psi} \right)(\vett{p}) =
\sum_{s = \pm 1}
1_{\mathcal{M}_s}(\theta) \, \tilde{\vett{e}}_s(\vett{p})  \,
\Big( \tilde{\vett{e}}^*_s(\vett{p}) \cdot \bgreek{\psi}(\vett{p}) \Big)
\end{equation}
preserves the idempotence property which characterizes a PVM, i.e., $\hat{F}_{S'_z, \epsilon}^2(\mathcal{M}) = \hat{F}_{S'_z, \epsilon}(\mathcal{M})$, and therefore the probability distribution of $\hat{S}'_z$ reads:
\begin{equation}
\label{eq:ddpsprime}
p_{S'_z}(\mathcal{M}) = \sum_{s=\pm 1}
\int dp \,\, p \,\, \int_{\mathcal{M}_s} d\theta \, \sin(\theta) \,\, \int_0^{2\pi} d\phi \,\,
|\tilde{\vett{e}}^*_s(\vett{p}) \cdot \bgreek{\psi}(\vett{p})|^2
\end{equation}
We remark that since $\hsys$ is left invariant by the projector $\YUPPI$, \eqref{eq:POVMsprime} is a PVM:
the SAM relative to the non-canonical decomposition~\eqref{eq:dec_alternatA}, \eqref{eq:dec_alternatB} is thus a {\em{sharp quantum observable}}.
Moreover, unlike its counterpart in the canonical decomposition, it can take all possible values inside the interval $[-1,1]$ \cite{Bliokh2012}.

Let us now consider the OAM $\hat{L}'_z$. Its action onto a function $\vett{f}(\vett{p}) \in \h_A$ reads:
\begin{equation}
\label{eq:lprimeeiva}
\left( \hat{L}'_z \, \vett{f} \, \right)(\vett{p}) =
- i \hbar \, \partial_\phi \, \vett{f}(\vett{p}) + H(\theta,\phi) \,
\vett{f}(\vett{p})
\quad , \quad \quad
H(\theta,\phi) = - \frac{1}{|\vett{p}|^2} \,
\left( \vett{p} \times (\vett{p} \times \vett{S}) \right)_z
\end{equation}
On the extended Hilbert space $\hbarra_A$, the generalized eigenfunctions of $\hat{L}'_z$ are readily worked out starting from~\eqref{eq:lprimeeiva}, details are reported in the Appendix~\ref{eq:eifunLprime}. They read
\begin{equation}
\vett{u}_{n,j}(\vett{p}) = f(|\vett{p}|) \, \delta(\cos(\theta)- s_0) \,
\vett{v}_{n,j}(\theta,\phi)
\end{equation}
where $s_0 \in [-1,1]$, $f(|\vett{p}|)$ is a properly normalized function, $\vett{v}_{n,j}(\theta,\phi)$ is detailed in Appendix~\ref{eq:eifunLprime} and $\hat{L}'_z \, \vett{u}_{n,j}(\vett{p}) = \hbar \, (j \, s_0 + n) \, \vett{u}_{n,j}(\vett{p})$.

The choice $j=0$ leads to unphysical eigenfunctions such that $\hat{\YUPPI} \,  \hat{V}^\dag \,  \vett{u}_{n,0}(\vett{p}) = 0$ (i.e. $\hat{V}^\dag \,  \vett{u}_{n,0}(\vett{p})$ is a longitudinal function).
On the other hand, the choices $j=\pm 1$ produce eigenfunctions such that $\hat{\YUPPI} \, \hat{V}^\dag \, \vett{u}_{n,0}(\vett{p}) = \vett{u}_{n,0}(\vett{p})$, i.e. $\hat{V}^\dag \vett{u}_{n,0}(\vett{p}) \in \hsys$.
In the light of these observations, the PVM associated to the OAM observable
on $\hbarra_A$ is:
\begin{equation}
\label{eq:PVMlprime}
\mathcal{M} \mapsto \left( \hat{E}_{L'_z}(\mathcal{M}) \, \vett{f} \right)(\vett{p}) = \sum_{j=\pm 1} \sum_{n\in\mathbb{Z}}
\, 1_{\mathcal{M}_{n,j}}(\theta) \,
\vett{v}_{n,j}(\theta,\phi)  \,
\int_0^{2\pi} d\phi \,
\Big( \vett{v}^*_{n,j}(\theta,\phi) \cdot \vett{f}(|\vett{p}|,\theta,\phi) \Big)
\end{equation}
where $\mathcal{M}$ is a Borel subset of $\mathbb{R}$ and $\mathcal{M}_{n,j}$ is
the set of angles $\theta$ such that $j \cos(\theta) + n \in \mathcal{M}$.
The corresponding POVM on $\hsys$ reads
\begin{equation}
\label{eq:POVMlprime}
\mathcal{M} \mapsto \left( \hat{F}_{L'_z}(\mathcal{M}) \, \bgreek{\psi} \right)(\vett{p}) = \sum_{j=\pm 1} \sum_{n\in\mathbb{Z}}
\, 1_{\mathcal{M}_{n,j}}(\theta) \,
\hat{V}^\dag \vett{v}_{n,j}(\theta,\phi)  \,
\int_0^{2\pi} d\phi \,
\Big( \vett{v}^*_{n,j}(\theta,\phi) \cdot \hat{V}\bgreek{\psi}(|\vett{p}|,\theta,\phi) \Big)
\end{equation}
and the probability distribution of $\hat{L}'_z$ is
\begin{equation}
\label{eq:ddplprime}
p_{L'_z}(\mathcal{M}) = \sum_{j=\pm 1} \sum_{n\in\mathbb{Z}}
\int_0^\infty dp \, p \, \int_{\mathcal{M}_{n,j}} d\theta \, \sin(\theta) \,
\left|
\int_0^{2\pi} d\phi \,
\Big( \vett{v}^*_{n,j}(\theta,\phi) \cdot \hat{V} \bgreek{\psi}(|\vett{p}|,\theta,\phi) \Big)\;
\right|^2.
\end{equation}
As in the case of the SAM observable, since $\hsys$ is left unchanged by the projectors~\eqref{eq:PVMlprime},~\eqref{eq:POVMlprime} is a PVM.


\subsection{Application to Gaussian states}\label{sec:Results}

For illustration, we show a direct application of the above formalism by explicitly calculating the first two cumulants (mean value and variance) of the probability distributions for both the canonical and the non-canonical SAM and OAM over \textit{circularly polarized} Gaussian single-photons,  i.e. wavefunctions $\bgreek{\psi}(\vett{p}) \in \h_{\mathcal{S}}$ of the form
\begin{equation}
\label{psi1}
\bgreek{\psi}(\vett{p}) =
\sqrt{|\vett{p}|} \, \frac{e^{- \frac{|\vett{p}-\vett{p}_0|^2}{8 a p_0^2} }}{\big( 4 \pi a p_0^2 \big)^{\frac{3}{4}}} \otimes \tilde{\vett{e}}_+(\vett{p}),
\end{equation}
where $\vett{p}_0 = p_0 \, \vett{e}_z$ and
\begin{equation}
a = \frac{(\Delta p)^2}{2 p_0^2}
\end{equation}
denoting a positive, dimensionless parameter which takes into account the spread of the wavefunction in momentum space.
We stress that, while any of the three components $x,y,z$ of these operators can be in principle evaluated, we will show the result for the $z$-component of the observables $\mathbf{L}, \mathbf{S}, \mathbf{L'}$ and $\mathbf{S'}$, i.e. the one parallel to the chosen $\vett{p_0}$.

\begin{figure*}[htbp!]
\begin{center}
\begin{tikzpicture}
  \node (img1)  {\includegraphics[scale=0.6]{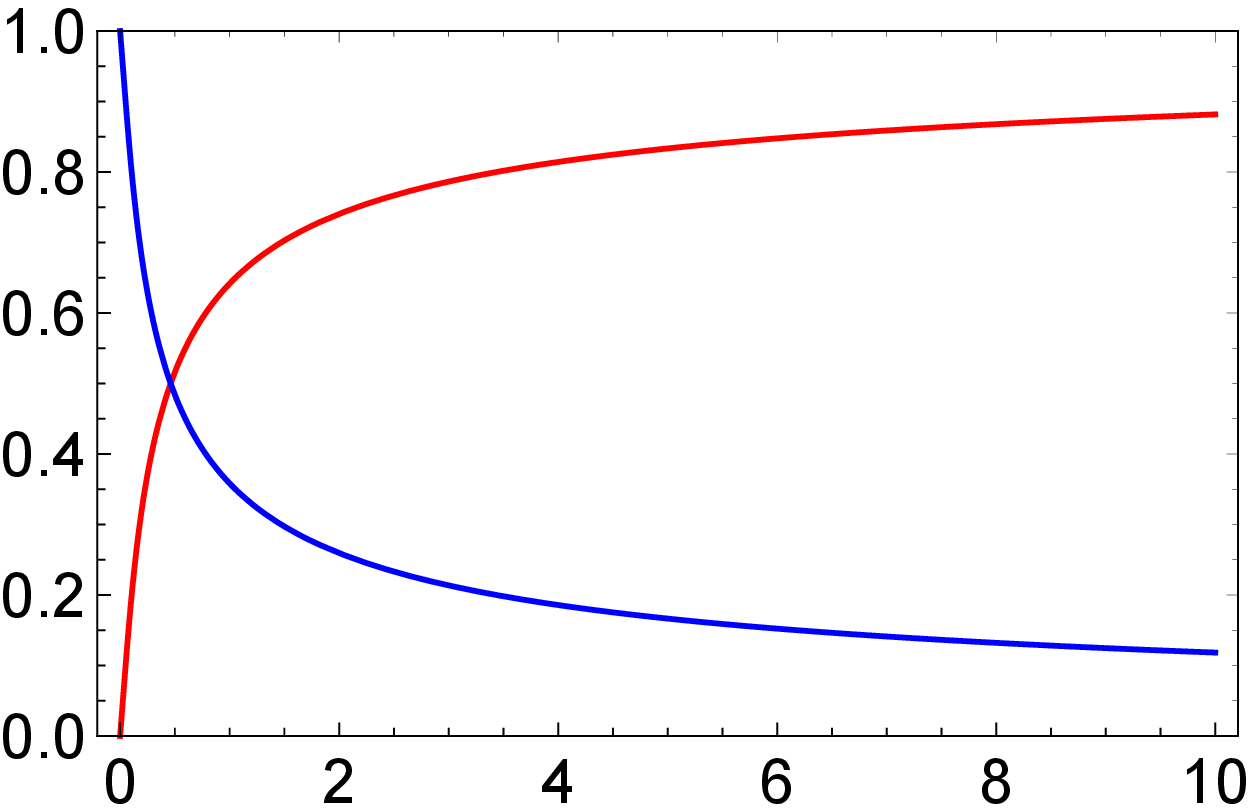}};
    \node[above=of img1, node distance=0cm, yshift=-2.6cm,xshift=1.5cm] {{\color{red}$\mean{L_z} = \mean{L'_z}$}};
    \node[above=of img1, node distance=0cm, yshift=-1.7cm,xshift=3.5cm] {{\color{black}{\bf{(a)}}}};
 \node[above=of img1, node distance=0cm, yshift=-4.8cm,xshift=1.5cm] {{\color{blue}$\mean{S_z} = \mean{S'_z}$}};
       \node[above=of img1, node distance=0cm, yshift=-6.4cm,xshift=3cm] {{\fontsize{14}{18}$a$}};
\end{tikzpicture}
\hspace*{0.5cm}
\begin{tikzpicture}
\node (img2)  {\includegraphics[scale=0.6]{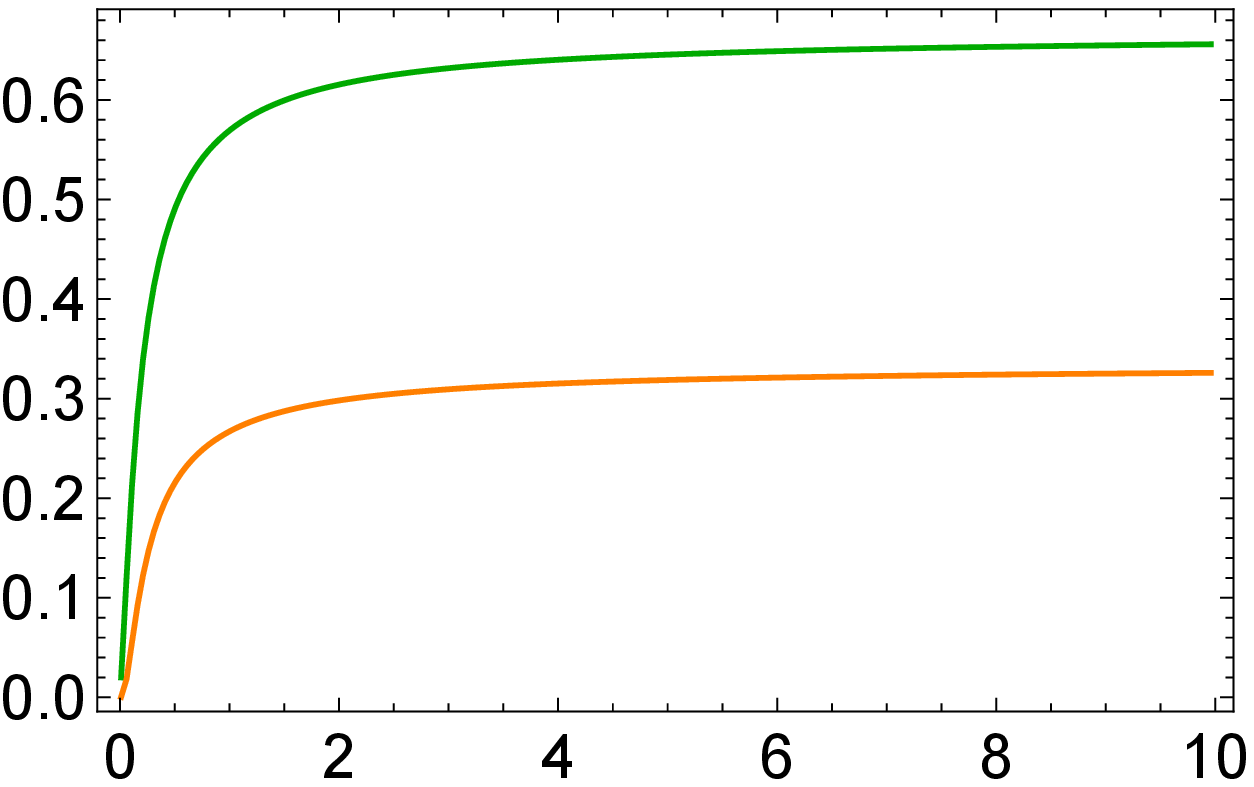}};
   \node[above=of img1, node distance=0cm, yshift=-2.2cm,xshift=1.5cm] {{\color{black!40!green}$\mathrm{Var}(L_z) = \mathrm{Var}(S_z)$}};
      \node[above=of img1, node distance=0cm, yshift=-1.9cm,xshift=3.3cm] {{\color{black}{\bf{(b)}}}};
 \node[above=of img1, node distance=0cm, yshift=-4.2cm,xshift=1.5cm] {{\color{black!20!orange}$\mathrm{Var}(L'_z) = \mathrm{Var}(S'_z)$}};
       \node[above=of img1, node distance=0cm, yshift=-6.4cm,xshift=3cm] {{\fontsize{14}{18}$a$}};
\end{tikzpicture}
\caption{Dependence of the first two cumulants of the two decompositions of the TAM (canonical and non-canonical) with respect to the parameter $a = \frac{(\Delta p)^2}{2 p_0^2}$ that quantifies the spread of the wavefunction in the momentum space. {\bf{(a)}} Plots of the mean values of the $z-$ components of $\mathbf{L}, \mathbf{S}, \mathbf{L'}$ and $\mathbf{S'}$; {\bf{(b)}} Plots of the respective variances.}
\label{Fig1}
\end{center}
\end{figure*}

Remarkably, the mean values of the two decompositions of the TAM are equal to each other on this particular state, plotted also in Fig.~\ref{Fig1}~{\bf{(a)}},
\begin{equation}
\mean{L_z} = \mean{L'_z} = 1 - f(a), \qquad \qquad \mean{S_z} = \mean{S'_z} = f(a),
\end{equation}
where
\begin{equation}
f(a) \equiv (1-2 a) \text{erf}\left(\frac{1}{2\sqrt{a}}\right) + \frac{2 \sqrt{a} e^{-\frac{1}{4a}}}{\sqrt{\pi}}.
\end{equation}

This shows that, for this particular class of single-photon states, there is no quantitative difference between the two decompositions of the total angular momentum at the level of the mean value. Moreover, the values of spin and momentum equal for $a=\frac{1}{2}$ corresponding to $\Delta p=p_0$, in this case both contributions to the TAM are maximal uncertain. For the paraxial limit $a\to 0^+$ the mean value of the TAM is dominated by the spin part, whereas for increasing $a$ the mean value of the TAM is basically the angular momentum, showing the dominant physical behaviour for high and low energetic photons (for fixed $\Delta p$), respectively. 

The departure of the two decompositions is instead witnessed at the level of the respective variances, shown in Fig.~\ref{Fig1}~{\bf{(b)}}, which contains all the crucial information about their statistical character.
In accordance with all the formal construction outlined above, the unsharpness of the canonical OAM and SAM brought by the introduction of the POVMs is in fact reflected in a larger value of the variance with respect to the (sharp) non-canonical decomposition for every value of the parameter $a$.
One can finally notice that, in the paraxial limit $a\to 0^+$, the two groups of variances correctly vanish, as the wavefunction is by construction perfectly defined in the momentum space.

\section{Conclusions}
\label{sec:conclusions}

We have studied in detail two different separations of the single-photon angular momentum into a spin and an orbital part, relying on the generalization~\cite{GML2015JphysA} of Karl Kraus' construction of the position observable.
The \textit{canonical} decomposition of the total angular momentum (TAM) into the spin angular momentum (SAM) and the orbital angular momentum (OAM) are
both compatible observables, however, have to be described by POVMs (positive-operator valued measures)
due to their incompatibility with the transversality condition.
The \textit{non-canonical} decomposition of the TAM proposed firstly by S. J. van Enk and G Nienhuis \cite{VanEnk1994} guarantees the compatibility with the transversality condition by paying the prize that the spatial and internal spin degrees of freedom get coupled. Thanks to our unified and general framework for dealing with generic single-photon observables, we could deduce the form of the non-canonical OAM and SAM as projector-valued measurements (PVMs). This proves how the transversality condition categorizes into principally unsharp or sharp observables.

Last but not least we have quantitatively shown the difference between the two above-mentioned decompositions of the TAM by calculating the first two cumulants (mean values and variances) on circularly polarized Gaussian states. The results allowed to clearly emphasize the \textit{unsharp} character of the canonical OAM and SAM, in contrast with the \textit{sharpness} of the non-canonical OAM and SAM, as is reflected by their larger variances. Moreover, it shows that independently of the decomposition polarisation becomes a well defined property for low energetic photons since the mean $z$-component becomes small.

This work could pave a new unified framework to handle also more photons in a quantum information theoretic way.

{\bf Acknowledgements --}  G.G. acknowledges the support of the Czech Science Foundation (GACR) (grant No. GB14-36681G). B.C.H. acknowledges gratefully the Austrian Science Fund (FWF-P26783).

\bibliographystyle{ieeetr}

\begin{thebibliography}{99}
\bibitem{Allen1992} L. Allen, M. W. Beijersbergen, R. J. C. Spreeuw and J. P. Woederman,
{\em{Phys. Rev. A}} {\bf{45}}, 8185 (1992)
\bibitem{Molina2007} G. Molina-Terriza, J. P. Torres and L. Torner, {\em{Nature Physics}} {\bf{3}},
305 (2007)
\bibitem{Nagali2009} E. Nagali, F. Sciarrino, F. De Martini, L. Marrucci, B. Piccirillo, E. Karimi
and E. Santamato {\em{Phys. Rev. Lett.}} {\bf{103}}, 013601 (2009)
\bibitem{Krenn2014} M. Krenn et al., {\em{N. J. Phys.}} {\bf{16}} (2014)
\bibitem{Hiesmayr2016} B.C. Hiesmayr, M.J.A. de Dood and W. L offler, Phys. Rev. Lett. {\bf 116}, 073601 (2016)
\bibitem{Hiesmayr2017} G. Carvacho, F. Graffitti, V. D'Ambrosio, B. C. Hiesmayr and F. Sciarrino,
Scientific Reports {\bf 7}, 13265 (2017)
\bibitem{Freedman1972} S. J. Freedman and J. F. Clauser, {\em{Phys. Rev. Lett.}} {\bf{28}}, 938 (1972)
\bibitem{Ou1988} Z. Y. Ou and L. Mandel, {\em{Phys. Rev. Lett.}} {\bf{61}}, 50 (1988)
\bibitem{Shih1988} Y. H. Shih and C. O. Alley, {\em{Phys. Rev. Lett.}} {\bf{61}}, 2921 (1988)
\bibitem{Tapster1994} P. R. Tapster, J. G. Rarity and P. C. M. Owens {\em{Phys. Rev. Lett.}} {\bf{73}}, 1923 (1994)
\bibitem{Rowe2001} M. A. Rowe, D. Kielpinski, V. Meyer, C. A. Sackett, W. M. Itano, C. Monroe and
D. J. Wineland, {\em{Nature }} {\bf{409}}, 791 (2001)
\bibitem{Kocsis2011} S. Kocsis, B. Braverman, S. Ravets, M. J. Stevens, R. P. Mirin,
L. Krister Shalm and A. M. Steinberg, {\em{Science}} {\bf{332}}, 1170 (2011)
\bibitem{Shadbolt2014} P. Shadbolt, J. C. F. Mathews, A. Laing and J. L. O'Brien,
{\em{Nature Physics}} {\bf{10}}, 278 (2014)
\bibitem{Humblet1943} J. Humblet, {\em{Physics}} {\bf{10}}, 585 (1943)
\bibitem{Franke2008} S. Franke-Arnoldl, L. Allen an M. J. Padgett, {\em{Laser Photon. Rev.}} {\bf{2}} 299 (2008)
\bibitem{Akhiezer1965} A. I. Akhiezer and V. B. Berestetskii, {\em{Quantum Electrodynamics}}, John Wiley \& Sons, New York (1965)
\bibitem{VanEnk1994} S. J. van Enk and G. Nienhuis, {\em{Europhys. Lett.}} {\bf{25}}, 497 (1994), {\em{J. Mod. Opt.}} {\bf{41}}, 963 (1994)
\bibitem{Barnett1994} S.M. Barnett and L. Allen, {\em{Opt. Commun.}} {\bf{110}}, 670 (1994)
\bibitem{Barnett2010} S.M. Barnett, {\em{J. Mod. Opt.}} {\bf{57}}, 1339 (2010)
\bibitem{Bliokh2010} K. Y. Bliokh, M. A. Alonso, E. A. Ostrovskaya and A. Aiello, {\em{Phys. Rev. A}} {\bf{82}}, 063825 (2010)
\bibitem{Bliokh2012} K. Y. Bliokh, F. Nori, Phys. Rev. A \textbf{86}, 033824 (2012)
\bibitem{Berestetskii1982} E. M. Lifshitz, L. P. Pitaevskii and V. B. Berestetskii, {\em{Quantum Electrodynamics}}, Butterworth-Heineman, Oxford (1982)
\bibitem{Soper1976} D. E. Soper, {\em{Classical Field Theory}}, John Wiley \& Sons, New York (1976)
\bibitem{Jackson1998} J. D. Jackson, {\em{Classical Electrodynamics}}, John Wiley \& Sons, New York (1998)
\bibitem{GML2015JphysA} G. Guarnieri, M. Motta and L. Lanz, {\em{J. Phys. A: Math. Theor.}} {\bf{48}} 265302 (2015)
\bibitem{Kraus1977} K. Kraus, {\em{Position Observable for the Photon}} in {\em{Uncertainty Principles and the Foundation of Quantum Mechanics}},
                    W. C. Price and S. Chissich ed, John Wiley \& Sons, New York (1977)
\bibitem{Lahti1995} P. Busch, M. Grabowski and P. Lahti, {\em{Operational Quantum Physics}}, Lecture Notes in Physics Monographs, Springer (1995)
\bibitem{Holevo1982} A. S. Holevo, {\em{Probabilistic and statistical aspects of quantum theory}}, North-Holland Publ. Cy., Amsterdam (1982)
\bibitem{Moses1967} H.E. Moses, {\em{J. Math. Phys.}} \textbf{8}, 1134 (1967)
\bibitem{Moses1968} H.E. Moses, {\em{J. Math. Phys.}} \textbf{9}, 16 (1968)
\bibitem{Wigner1939} E. P. Wigner, Ann. Math. \textbf{40}, 149 (1939)
\bibitem{Hamermesh1989} M. Hamermesh, {\em{Group theory and its application to physical problems}}, Dover Publications (1989)
\bibitem{Busch1996} P. Busch and A. Shimony, {\em{Stud. Hist. Phil. Mod. Phys.}} {\bf{27}}, 397 (1996)
\bibitem{Busch2009} P. Busch, {\em{Found. Phys.}} {\bf{39}}, 712 (2009)
\bibitem{Busch2010} P. Busch and G. Jaeger, {\em{Found. Phys.}} {\bf{40}}, 1341 (2010)
\bibitem{Hawton2010} M. Hawton {\em{Phys. Rev. A}} {\bf{82}}, 012117 (2010)
\bibitem{Mandel1995} L. Mandel and E. Wolf, {\em{Optical Coherence and Quantum Optics}}, Cambridge University Press, Cambridge (1995)
\bibitem{Berry2009} M. V. Berry,  International Society for Optics and Photonics,  p. 6-12 (1998)
\bibitem{Li2009} C. F. Li, {\em{Phys. Rev. A}} {\bf{80}} 063814 (2009)
\bibitem{Hawton1999} M. Hawton, {\em{Phys. Rev. A}} {\bf{59}} 954 (1999)

\bibitem{Robertson}
H.P. Robertson, Phys. Rev. {\bf 34}, 163 (1929)

\bibitem{entropicuncertainty}
I. Bialynici-Birula and L. Rudnicki,
Statistical Complexity, Ed. K. D. Sen, Springer, Ch. 1 (2011) 

\bibitem{entropicuncertainty2}
A. Di Domenico, A. Gabriel, B.C. Hiesmayr,  et al., Found Phys {\bf 42}, 778 (2012)

\bibitem{Massar2007} S. Massar, {\em{Phys. Rev. A}} {\bf{76}} 042114 (2007)
\bibitem{Bliokh2014} K. Y. Bliokh, J. Dressel and F. Nori, {\em{New J. Phys.}} {\bf{16}}, 093037 (2014)
\bibitem{NoriReview} K. Y. Bliokh and F. Nori, {\em{Phys. Rep.}} {\bf{592}}, 1-38 (2015)


\end{thebibliography}

\newpage
\appendix

\section{Commutation relations for the canonical SAM and OAM}
\label{app:commutators}

In this appendix we show the commutation relations first for the canonical decomposition of the TAM, the for the non-canonical decomposition. In the last part we give the details on the derivation of the eigenfunctions of $\hat{L}'_z$.

We start by computing the commutators of the SAM and OAM operators and deriving the TAM through the canonical decomposition; to explicitly show that their incompatibility with the transversality condition traduces in a violation of Eq.~\eqref{pvm2povm_caveat}, i.e., in a non-vanishing commutator with the projection from the extended Hilbert space $\hbarra_{\mathcal{A}}$ onto the physical one $\hsys$.

Denoting with $\hat{A}_k$ the generators of the $so(3)$ algebra, i.e. the three antisymmetric matrices $(A_k)_{ml} = \epsilon_{kml}$,
by virtue of Eq.~\eqref{su2so3} we have that
\begin{equation}
\hat{V}^\dag \hat{S}_k \hat{V} = i \hbar \hat{A}_k\;.
\end{equation}
Then $ \left[ \hat{V}^\dag \hat{S}_k \hat{V}, \hat{\YUPPI} \right] = i \hbar \left[ \hat{A}_k , \hat{\YUPPI} \right] $. Now, since
\begin{equation}
\begin{split}
\left( \hat{A}_k \hat{\YUPPI} \bgreek{\psi}\right)_l (\vett{p}) &= \sum_{mr} (\hat{A}_k)_{lm} \hat{\YUPPI}(\vett{p})_{mr} \bgreek{\psi}_r(\vett{p}) =
\sum_{mr} \epsilon_{klm} \left(\delta_{mr} - \frac{p_m p_r}{|\vett{p}|^2} \right) \bgreek{\psi}_r(\vett{p}) = \\
&= (\hat{A}_k \bgreek{\psi}(\vett{p}))_l - \vett{p}\cdot\bgreek{\psi}(\vett{p}) \, \frac{(\hat{A}_k \vett{p})_l }{|\vett{p}|^2}
\end{split}
\end{equation}
and
\begin{equation}
\begin{split}
\left( \hat{\YUPPI} \hat{A}_k \bgreek{\psi}\right)_l (\vett{p}) &= \sum_{mr} \hat{\YUPPI}(\vett{p})_{lm} (\hat{A}_k)_{mr}  \bgreek{\psi}_r(\vett{p}) =
\sum_{mr} \left(\delta_{lm} - \frac{p_l p_m}{|\vett{p}|^2} \right) \epsilon_{kmr} \bgreek{\psi}_r(\vett{p}) = \\
&= (\hat{A}_k \bgreek{\psi}(\vett{p}))_l -
\frac{(\vett{p}\times\bgreek{\psi}(\vett{p}))_k}{|\vett{p}|^2} \, p_l
\end{split}
\end{equation}
it is immediately found that
\begin{equation}
\begin{split}
\left( [\hat{A}_k, \hat{\YUPPI}]\; \bgreek{\psi}\right)_l (\vett{p}) &= \frac{(\vett{p}\times\bgreek{\psi}(\vett{p}))_k}{|\vett{p}|^2} \, p_l
- \frac{(\hat{A}_k \vett{p})_l }{|\vett{p}|^2} \, \vett{p}\cdot\bgreek{\psi}(\vett{p})\;,
\end{split}
\end{equation}
which implies
\begin{equation}
\label{eq:comparison1}
\begin{split}
[\hat{V}^\dag \hat{S}_k \hat{V} , \hat{\YUPPI}]\; \bgreek{\psi} (\vett{p}) &= i \hbar \left(
\frac{(\vett{p}\times\bgreek{\psi}(\vett{p}))_k}{|\vett{p}|^2} \, \vett{p}
- \vett{p}\cdot\bgreek{\psi}(\vett{p}) \, \frac{\hat{A}_k \vett{p}}{|\vett{p}|^2}
\right).
\end{split}
\end{equation}
To retrieve the second of equations \eqref{eq:commutators_AM}, it must be observed that
\begin{equation}
\left( \hat{V}^\dag \hat{L}_k \hat{V} \bgreek{\psi} \right)_l(\vett{p}) = \sum_{mr} \hat{V}^\dag_{lm}
( - i \hbar \, \partial_{\vett{p}} \times \vett{p} )_k \hat{V}_{mr} \bgreek{\psi}_r(\vett{p})
= \left( \hat{L}_k \bgreek{\psi} \right)_l(\vett{p})
\end{equation}
recalling the unitarity of $\hat{V}$. Moreover, since
\begin{equation}
\begin{split}
\left( \hat{L}_k \hat{\YUPPI} \bgreek{\psi} \right)_l(\vett{p}) &= \sum_{r}
( - i \hbar \, \partial_{\vett{p}} \times \vett{p} )_k \left( \hat{\YUPPI}_{lr}(\vett{p})  \bgreek{\psi}_r(\vett{p}) \right) = \\
&= \sum_r ( - i \hbar \, \partial_{\vett{p}} \times \vett{p} )_k \left( \hat{\YUPPI}_{lr}(\vett{p}) \right) \bgreek{\psi}_r(\vett{p})
+ \sum_r \hat{\YUPPI}_{lr}(\vett{p}) ( - i \hbar \, \partial_{\vett{p}} \times \vett{p} )_k \left( \bgreek{\psi}_r(\vett{p}) \right) = \\
&= \sum_r ( - i \hbar \, \partial_{\vett{p}} \times \vett{p} )_k \left( \hat{\YUPPI}_{lr}(\vett{p}) \right) \bgreek{\psi}_r(\vett{p})
+\left( \hat{\YUPPI} \hat{L}_k  \bgreek{\psi} \right)_l(\vett{p})
\end{split}
\end{equation}
one has
\begin{equation}
\left( [\hat{L}_k , \hat{\YUPPI}] \bgreek{\psi} \right)_l(\vett{p}) =
\sum_r ( - i \hbar \, \partial_{\vett{p}} \times \vett{p} )_k \left( \hat{\YUPPI}_{lr}(\vett{p}) \right) \bgreek{\psi}_r(\vett{p})
\end{equation}
and since
\begin{equation}
\begin{split}
( - i \hbar \, \partial_{\vett{p}} \times \vett{p} )_k \left( \hat{\YUPPI}_{lr}(\vett{p}) \right) &=
- i \hbar \sum_{ms} \epsilon_{ksm} p_m \partial_{p_s} \left( \delta_{lr} + \frac{p_l p_r}{|\vett{p}|^2} \right) = \\
&= i \hbar \sum_{ms} \epsilon_{ksm} p_m \frac{(\delta_{ls} p_r + \delta_{rs} p_l) |\vett{p}|^2 - 2 p_l p_r p_s}{|\vett{p}|^4} = \\
&= i \hbar \frac{p_r \sum_{m} \epsilon_{klm} p_m + p_l \sum_m \epsilon_{krm} p_m }{|\vett{p}|^2} = \\
&= i \hbar \frac{p_r (A_k \vett{p})_l + p_l (\hat{A}_k \vett{p})_r}{|\vett{p}|^2}
\end{split}
\end{equation}
the result
\begin{equation}
\label{eq:comparison2}
[\hat{L}_k , \hat{\YUPPI}] \bgreek{\psi}(\vett{p}) = i \hbar \left( \vett{p} \cdot \bgreek{\psi}(\vett{p}) \,
\frac{\hat{A}_k \vett{p}_l }{|\vett{p}|^2} -
\frac{\left( \vett{p} \times \bgreek{\psi}(\vett{p}) \right)_k}{|\vett{p}|^2} \, \vett{p} \right)\;.
\end{equation}
To summarize, the commutators of the SAM and OAM with the projection onto the physical Hilbert space do not vanish and are equal and opposite to each other:
\begin{align}\label{eq:commutators_AM}
\left[ \hat{V}^\dag \hat{S}_k \hat{V} , \hat{\YUPPI} \right] \vett{f}(\vett{p}) &= i \hbar
   \left( \frac{\left( \vett{p} \times \vett{f}(\vett{p}) \right)_k}{|\vett{p}|^2} \, \vett{p}
- \vett{p} \cdot \vett{f}(\vett{p}) \, \frac{A_k \vett{p}}{|\vett{p}|^2} \right), \\
\left[ \hat{V}^\dag \hat{L}_k \hat{V} , \hat{\YUPPI} \right] \vett{f}(\vett{p}) &= -
\left[ \hat{V}^\dag \hat{S}_k \hat{V} , \hat{\YUPPI} \right] \vett{f}(\vett{p}).
\end{align}
An immediate consequence of the last result is that the commutator of the TAM, $\hat{J}_k = \hat{L}_k+\hat{S}_k$, with the projector $\hat{V}\hat{\YUPPI}\hat{V}^\dag$ is zero and thus  we retrieve the result that the total angular momentum is a sharp observable for single photons, i.e. is consistent with the transversality condition and thus is described in terms of a PVM.

\section{Commutation relations for the non-canonical SAM and OAM}

We now prove that $[\hat{S}'_i,\hat{L}'_j] \neq 0$, and $[\hat{L}'_i,\hat{L}'_j] \neq 0$. To prove the first property, it is sufficient to write:
\begin{equation}
\hat{S}^\prime_k = \sum_{l} \frac{ p_k p_l }{|\vett{p}|^2} \, \hat{S}_l
\quad
\quad , \quad
\hat{L}^\prime_k = \hat{L}_k + \hat{S}_k - \hat{S}^\prime_k
\end{equation}
whence:
\begin{equation}
\label{eq:commLpSp}
[ \hat{L}^\prime_i , \hat{S}^\prime_j] =
\sum_l \left[ \hat{L}_i , \frac{ p_j p_l }{|\vett{p}|^2} \, \hat{S}_l \right]
+
\sum_l \, \frac{p_j p_l}{|\vett{p}|^2} \, [\hat{S}_i , \hat{S}_l]
- [\hat{S}^\prime_i , \hat{S}^\prime_j] =
\sum_l \left[ \hat{L}_i , \frac{ p_j p_l }{|\vett{p}|^2} \right] \, \hat{S}_l
+ \sum_l \, \frac{p_j p_l}{|\vett{p}|^2} \, [\hat{S}_i , \hat{S}_l]
\end{equation}
The second term of the sum \eqref{eq:commLpSp} can be easily worked out
recalling the commutation properties of Pauli matrices:
\begin{equation}
\sum_l \frac{p_j p_l}{|\vett{p}|^2} \,\, [\hat{S}_i , \hat{S}_l] =
\sum_{lr} \frac{p_j p_l}{|\vett{p}|^2} \,\, i \hbar \,\, \epsilon_{ilr} \,\, \hat{S}_r =
i \hbar \,\, \frac{p_j}{|\vett{p}|} \, \left( \frac{\vett{p}}{|\vett{p}|} \times \hat{\vett{S}} \right)_i
\end{equation}
while the first term reads:
\begin{equation}
\sum_l \left[ \hat{L}_i , \frac{ p_j p_l }{|\vett{p}|^2} \right] \, \hat{S}_l =
- i \hbar \,
\left( \sum_k \epsilon_{ijk} \, \frac{p_k}{|\vett{p}|} \right) \, \frac{\vett{p} \cdot \hat{\vett{S}}}{|\vett{p}|} - i \hbar \, \frac{p_i}{|\vett{p}|} \, \left( \frac{\vett{p}}{|\vett{p}|} \times \hat{\vett{S}} \right)_j
\end{equation}
In conclusion:
\begin{equation}
\label{eq:commlpsp2}
[ \hat{L}^\prime_i , \hat{S}^\prime_j] =
i \hbar \, \frac{p_j}{|\vett{p}|} \, \left( \frac{\vett{p}}{|\vett{p}|} \times \hat{\vett{S}} \right)_i
- i \hbar \, \frac{p_i}{|\vett{p}|} \, \left( \frac{\vett{p}}{|\vett{p}|} \times \hat{\vett{S}} \right)_j
- i \hbar \,
\left( \sum_k \epsilon_{ijk} \, \frac{p_k}{|\vett{p}|} \right) \, \frac{\vett{p} \cdot \hat{\vett{S}}}{|\vett{p}|}
\end{equation}
Eq.~ \eqref{eq:commlpsp2} also shows that the components of the OAM operator $\hat{L}'_k$ do not commute with each other:
\begin{equation}
\label{eq:commlpsp3}
[ \hat{L}^\prime_i , \hat{L}^\prime_j ] =
[ \hat{J}^\prime_i , \hat{J}^\prime_j ] -
[ \hat{S}^\prime_i , \hat{L}^\prime_j ] -
[ \hat{L}^\prime_i , \hat{S}^\prime_j ] \neq 0
\end{equation}

\section{Eigenfunctions of $\hat{L}'_z$}
\label{eq:eifunLprime}

The action of $\hat{L}'_z$ onto a function $\vett{f}(\vett{p}) \in \h_A$ reads:
\begin{equation}
\label{eq:lprimeeiva2}
\left( \hat{L}'_z \, \vett{f} \right)(\vett{p}) =
- i \hbar \, \partial_\phi \, \vett{f}(\vett{p}) + H(\theta,\phi) \,
\vett{f}(\vett{p})
\end{equation}
where the $3 \times 3$ matrix $H(\theta,\phi)$ reads:
\begin{equation}
H(\theta,\phi) =
- \frac{\left( \vett{p} \times (\vett{p} \times S) \right)_z}{|\vett{p}|^2} =
\begin{pmatrix}
-\sin ^2(\theta ) & \frac{e^{-i \phi } \sin (\theta ) \cos (\theta )}{\sqrt{2}} & 0 \\
 \frac{e^{i \phi } \sin (\theta ) \cos (\theta )}{\sqrt{2}} & 0 & \frac{e^{-i \phi } \sin
   (\theta ) \cos (\theta )}{\sqrt{2}} \\
 0 & \frac{e^{i \phi } \sin (\theta ) \cos (\theta )}{\sqrt{2}} & \sin ^2(\theta ) \\
\end{pmatrix}
\end{equation}
The eigenvalues of $H(\theta,\phi)$ are $\lambda_s = s \, \sin(\theta)$, $s= 0,\pm 1$, remarkably independent of $\phi$; the corresponding eigenvectors are the following periodic functions $\vett{v}_s(\theta,\phi)$ of $\phi$, with period $2\pi$:
\begin{align}
&\vett{v}_1(\theta,\phi) =
\begin{pmatrix}
 -\frac{1}{2} e^{-2 i \phi } \sec ^2(\theta ) (4 \sin (\theta )+\cos (2 \theta )-3) \\
 \sqrt{2} e^{-i \phi } (\sec (\theta )-\tan (\theta )) \\
 1 \\
\end{pmatrix}
\\
&\vett{v}_0(\theta,\phi) =
\begin{pmatrix}
-e^{-2 i \phi } \\
 -\sqrt{2} e^{-i \phi } \tan (\theta ) \\
 1 \\
\end{pmatrix}
\\
&\vett{v}_{-1}(\theta,\phi) =
\begin{pmatrix}
-\frac{1}{2} e^{-2 i \phi } \sec ^2(\theta ) (-4 \sin (\theta )+\cos (2 \theta )-3) \\
 -\sqrt{2} e^{-i \phi } (\tan (\theta )+\sec (\theta )) \\
 1\\
\end{pmatrix}
\end{align}
For all $\theta$, the eigenfunctions of $\hat{L}'_z$ must have the form:
\begin{equation}
\label{eq:lprime1}
\vett{v}(\theta,\phi) = \sum_{s=0,\pm 1} \alpha_s(\theta,\phi)
\,
\vett{v}_s(\theta,\phi)
\end{equation}
of linear combinations of the vectors $\vett{v}_s(\theta,\phi)$ with coefficients
$\alpha_s(\theta,\phi)$ that are periodic functions of $\phi$, with period $2\pi$. Inserting \eqref{eq:lprime1} in the eigenvalue equation for \eqref{eq:lprimeeiva2} we are led to the following eigenvalue equation:
\begin{equation}
- i \hbar \partial_\phi \alpha(\theta,\phi) - i G(\theta) \alpha(\theta,\phi) + \Lambda(\theta) \alpha(\theta,\phi) = \hbar m \, \alpha(\theta,\phi)
\quad, \quad\quad
\alpha(\theta,\phi) =
\begin{pmatrix}
\alpha_{-1}(\theta,\phi) \\
\alpha_0(\theta,\phi) \\
\alpha_1(\theta,\phi)
\end{pmatrix},
\end{equation}
where $\Lambda(\theta)_{rs} = \lambda_s(\theta) \, \delta_{rs}$,
\begin{equation}
G(\theta)_{rs} = \vett{v}_r^*(\theta,\phi) \partial_\phi \vett{v}_s(\theta,\phi) =
\begin{pmatrix}
 -i (\sin (\theta )+1) & \frac{i \cos (\theta )}{\sqrt{2}} & 0 \\
 \frac{i \cos (\theta )}{\sqrt{2}} & -i & \frac{i \cos (\theta )}{\sqrt{2}} \\
 0 & \frac{i \cos (\theta )}{\sqrt{2}} & i (\sin (\theta )-1) \\
\end{pmatrix}
\end{equation}
and the eigenvalue $m$ will be determined in a short while. Since the matrix $M(\theta) = -i G(\theta) + \Lambda(\theta)$, is remarkably independent of $\phi$, we find:
\begin{equation}
\alpha(\theta,\phi) = e^{- i M(\theta) \phi}
e^{i \left( \mu_j(\theta) + n \right) \phi} \vett{m}_j(\theta)
\end{equation}
where $M(\theta) \vett{m}_j(\theta) = \mu_j(\theta) \vett{m}_j(\theta)$ has eigenvalues $\mu_j(\theta) = 1 + j\cos\left(\theta\right)$ and eigenvectors $\vett{m}_j(\theta)$ explicitly given by
\begin{equation}
\vett{m}_1(\theta) =
\begin{pmatrix}
1 \\
-\sqrt{2} \\
1 \\
\end{pmatrix}
,\,\,
\vett{m}_0(\theta) =
\begin{pmatrix}
-1 \\
0 \\
1 \\
\end{pmatrix}
,\,\,
\vett{m}_{-1}(\theta) =
\begin{pmatrix}
1 \\
\sqrt{2} \\
1 \\
\end{pmatrix}
\end{equation}
Combining these results, we obtain the eigenfunctions of the non-canonical OAM $\hat{L}'_z$
\begin{equation}
\label{eq:lprime2}
\vett{v}_{n,j}(\theta,\phi) = \sum_s e^{- i n \phi} \left( \vett{m}_j \right)_s \,
\vett{v}_{s}(\theta,\phi)
\end{equation}
relative to the eigenvalues $\mu_j(\theta) + n$.

\end{document}